\documentclass[10pt,a4paper]{article}
\pdfoutput=1

\usepackage{graphicx}
\usepackage{amssymb, latexsym}
\usepackage{multirow}
\usepackage{amsmath}
\usepackage{mathrsfs}  
\usepackage{xcolor}
\usepackage[T1]{fontenc}
\usepackage[utf8]{inputenc}
\usepackage{authblk}
\usepackage{hyperref}

\usepackage[a4paper,top=2.5cm,bottom=2cm,left=1.8cm,right=1.8cm]{geometry}

\title{Investigating two heavy neutral leptons neutrino seesaw mechanism at SHiP
}

\author[1]{Marco Chianese\thanks{ma.chianese@gmail.com}}
\author[2,3]{Damiano F. G. Fiorillo\thanks{damianofg@gmail.com}}
\author[2,3]{Gennaro Miele\thanks{miele@na.infn.it}}
\author[2,3]{Stefano Morisi \thanks{stefano.morisi@gmail.com}}

\affil[1]{{\it \small Gravitation Astroparticle Physics Amsterdam (GRAPPA), Institute of Physics, University of Amsterdam, Science Park 904, 1098 XH Amsterdam, The Netherlands}}

\affil[2]{{\it \small INFN - Sezione di Napoli, Complesso Univ. Monte S. Angelo, I-80126 Napoli, Italy}}

\affil[3]{{\it \small Dipartimento di Fisica {\it "Ettore Pancini"}, Universit\`a degli studi di Napoli "Federico II", Complesso Univ. Monte S. Angelo, I-80126 Napoli, Italy}}

\begin{document}

\maketitle

\begin{abstract}
One of the main purposes of SHiP experiment is to shed light on neutrino mass generation mechanisms like the so-called seesaw.  We consider a minimal type-I seesaw neutrino mass mechanism model with two heavy neutral leptons (right-handed or sterile neutrinos) with arbitrary masses. Extremely high active-sterile mixing angle requires a correlation between the phases of the Dirac neutrino couplings. Actual experimental limits on the half-life of neutrinoless double beta decay $0\nu\beta\beta$-rate on the active-sterile mixing angle are not significative for SHiP.
\end{abstract}

\tableofcontents

\section{Introduction}

The experimental detection of neutrino mass is one of the most important evidences that new physics beyond the Standard Model is required. In the Standard Model neutrinos can get a Majorana mass by means of the non-renormalizable dimension five Weinberg operator \cite{Weinberg:1979sa} 
\begin{equation}
\frac{1}{\Lambda} \overline{L} \tilde{H}\overline{L} \tilde{H} \,,
\label{eq:Weinberg}
\end{equation}
where $\Lambda$ is some effective scale, $L$ and $H$ are respectively  the lepton and Higgs $SU(2)_L$ doublets  where $\tilde{H}=-i\sigma_2H^*$. Renormalization principle suggests that behind Weinberg operator it exists some extension beyond the Standard Model, called neutrino mass mechanism. Moreover, the possible underlying theory depends on the nature of neutrino fields, namely if they are Majorana or Dirac particles. In both cases  many models have been investigated and they can be distinguished in models where only new fermions, or scalars, or scalars plus fermions are introduced beyond the standard matter content, and the Weinberg operator can be generated by means of three-level Feynman diagram or radiative one \cite{Ma:1998dn}. The most popular model for neutrino mass known as type-I seesaw extends the Standard Model by means of Heavy Neutral Leptons (HNL), singlets under the electroweak symmetry, that are the right-handed components of active left-handed neutrinos. Thus an important experimental goal in order to better understand the origin of neutrino mass, is the detection of HNLs like in SHiP experiment \cite{Bonivento:2013jag,Anelli:2015pba}. SHiP plans to study masses of about $(0.6\div 6.0)$\,GeV \cite{Graverini:2015dka,SHiP:2018xqw}. It is commonly argued \cite{Boucenna:2014zba}  that with type-I seesaw mechanisms there are no detectable direct production signatures at colliders nor lepton flavor violation processes  even if the HNL mass scale is as low as $\mathcal{O}(1)$\,GeV because in first approximation the active-sterile mixing is expected to be proportional to
\begin{equation}\label{naive}
\frac{\sqrt{\Delta m^2_{atm}}}{M_{\rm HNL}}\approx 5\cdot 10^{-11} \left(\frac{1{\mbox GeV}}{M_{\rm HNL}}\right)\,,
\end{equation}
that is very suppressed compared to SHiP sensitivity \cite{Graverini:2015dka,SHiP:2018xqw} except for symmetry protected scenarios\,\cite{Antusch:2015mia} or low-energy seesaw mechanisms like \cite{Mohapatra:1986aw,Mohapatra:1986bd}.

In this work we study a simple neutrino mass model with two Heavy Neutral Leptons and show that the naive expectation~\eqref{naive} is wrong being the upper limit of possible values of the active-sterile mixing much larger than the one reported in expression~\eqref{naive}. On the other hand, we find that the lower limit for the value of active-sterile mixing set by seesaw mechanism is approximately ten orders of magnitude smaller than SHiP sensitivity. Hence, the stronger constraints on small values of active-sterile mixing are provided by Big Bang Nucleosynthesis.

The paper is organized as follows. In Section 2, we review seesaw neutrino mass mechanisms with HNLs. We analytically study  some cases and we provide analytical expressions for the upper and lower active-sterile mixing  limits. In Section 3, we give the details of our numerical analysis, including also neutrinoless double beta decay constraint and constraints coming from collider and BBN data, and show why the naive expectation~\eqref{naive} is wrong for the upper limits. Then, in Section 4 we draw our conclusions.

\section{Two Heavy Neutral Leptons seesaw and SHiP}

As discussed in the Introduction, we know from neutrino physics that the Standard Model must be extended and a well studied class of models makes use of new fermions. The most popular model in particular exploits $n$ Heavy Neutral Leptons or sterile neutrinos $N_{i}$ with $i=1,..,n$ that are  singlets under the electroweak symmetry and are the right-handed components of active left-handed neutrinos $\nu_{L_\alpha}$ with $\alpha=e,\mu,\tau$. The fields $N_{i}$ are assumed to be Majorana particles. In addition to Dirac mass interactions $M_D$ (that is a $3\times n$ mass matrix) that couple left- and right-handed neutrino components like in the charged lepton sector, thanks to the Majorana nature of the new Heavy Neutral Leptons, we can have interacting couplings  between $N_{i}$ themselves $M_{R}$ (that is a $n\times n$ mass matrix). Such couplings violate the lepton number by two units. Thus, the whole neutrino mass matrix $M_\nu$ in the $(\nu_{L_\alpha},\,N_{i})$ basis is a $(3+n)\times (3+n)$ symmetric matrix
\begin{equation}\label{mnu}
M_\nu =\left[
\begin{tabular}{cc}
0 & $(M_D)_{3\times n}$\\
$(M_D)_{3\times n}^T$ & $(M_{R})_{n\times n}$
\end{tabular}
\right]\,.
\end{equation}
This matrix is diagonalized with a unitary $(3+n)\times (3+n)$ matrix $U$ such that 
\begin{equation}\label{diagonalization}
U^T M_\nu U \equiv {\rm diag} (m_{\nu1},m_{\nu2},m_{\nu3},M_{1},....M_n) \,.
\end{equation}
We note that the $3\times3$ sub-block of $U$ is the Pontecorvo–Maki–Nakagawa– Sakata (PMNS) matrix $U_{\rm PMNS}$~\cite{Tanabashi:2018oca} that is parametrized by the three mixing angles $\theta_{12}^\nu$, $\theta_{13}^\nu$, $\theta_{23}^\nu$ and three phases $\delta$, $\alpha$, $\beta$. The mixing angles and one of the phases are measured in neutrino oscillation experiments (for a global analysis see \cite{Capozzi:2018ubv,deSalas:2018bym,Esteban:2018azc}). If $M_D\ll M_{R}$, it is well known that the mass matrix~\eqref{mnu} can be block-diagonalized and the resulting ($3\times 3$)  light neutrino mass matrix is given by  
\begin{equation}\label{ss1}
m_\nu^{\rm light}\approx -M_D\frac{1}{M_{R}}M_D^T\,.
\end{equation}
Since in relation~\eqref{ss1} the heavier $M_{R}$ the lighter $m_\nu$, such a mechanism is called seesaw: the nice feature is that if $M_D$ is at electroweak scale $\left(100~{\rm GeV}\right)$ and $M_{R}$ close to the grand unification scale $\left(10^{14}~{\rm GeV}\right)$, then light neutrinos are miraculously at eV scale. The seesaw mechanism given by relation~\eqref{ss1} is referred to as type-I. Other realizations of the seesaw mechanism make use of a scalar electroweak triplet (type-II) or a fermion electroweak triplet (type-III). All these seesaw mechanisms are just possible realizations of the dimension five Weinberg operator~\eqref{eq:Weinberg}. Apart from seesaw, other renormalizable realizations of the Weinberg operator have been proposed like the radiative one. For a review of all these neutrino mass mechanisms see for instance~\cite{Boucenna:2014zba,Grimus:2006nb} and references therein.

It is well known that in order to fit the two measured  neutrino square mass differences
$\Delta m^2_{sol}\equiv m^2_{\nu2}-m^2_{\nu1}$ and $\Delta m^2_{atm}\equiv m^2_{\nu3}-m^2_{\nu1}$ by using the relation\,(\ref{ss1}), at least two right-handed neutrinos are necessary, namely $n \ge 2$. So the simplest seesaw model fitting neutrino oscillation data has two HNLs, namely $n=2$ and gives $m_{\nu1}=0$. In the present work, we focus on such a model hereafter referred to as 2HNL seesaw.

We compare the 2HNL seesaw with the so-called $\nu$-Minimal Standard Model ($\nu$MSM) \cite{Asaka:2005an} (see also \cite{Asaka:2005pn,Atre:2009rg,Mohapatra:2006gs,Shaposhnikov:2006xi,Gorbunov:2007ak,Drewes:2013gca,Canetti:2012vf,Caputo:2016ojx}) that is another extension of the Standard Model where three right-handed are assumed instead of just two. Within $\nu$MSM
 it is possible  to explain the Baryon Asymmetry of the Universe (BAU) assuming almost degenerate 
HNLs at $(0.1\div10)$\,GeV scale (by means of leptogenesis) and provide a Dark Matter candidate at keV scale. Even if $\nu$MSM is a very  ambitious and interesting approach, seesaw mechanism could be  unrelated to the origin of baryon asymmetry and Dark Matter problems (like in 2HNL seesaw) and a model-independent analysis of 2HNL seesaw detection with SHiP is the task of the present work. On the other hand, baryon asymmetry and Dark Matter could be explained in a common framework not necessarily related with neutrino physics like in the so-called Asymmetric Dark Matter scenario~\cite{Kaplan:2009ag}. 

Within the 2HNL seesaw the neutrino mass matrix $M_\nu$~\eqref{mnu} is a  $5\times 5 $ matrix and the corresponding  diagonalizing unitary matrix $U$ is also a $5\times 5 $ matrix. We observe that it is always possible by means of a change of basis to take the $M_{R}$ matrix diagonal with entries $M_{1}$ and $M_{2}$. Then it is useful to parametrize the Dirac neutrino mass matrix $M_D$ as a function of the physical observables, the neutrino square mass differences, the mixing angles and phases. This is given by the Casas-Ibarra parametrization \cite{Casas:2001sr} as
\begin{equation}\label{Cl1}
M_D=U_{\rm PMNS} \, \sqrt{m^{\rm diag}_\nu} \,R \, \sqrt{M^{\rm diag}_{ R}}\,,
\end{equation}
where $m^{\rm diag}_\nu$ is the diagonal mass matrix with the three active neutrino masses, $M^{\rm diag}_{R}$ is the corresponding matrix for the HNLs and $R$ is an arbitrary orthogonal $3\times 2$ matrix given by (for normal neutrino mass ordering)
\begin{align}\label{Rmat}
    R=\left(\begin{array}{cc}
         0& 0  \\
         \cos\theta & \sin\theta \\
         -\kappa \sin\theta & \kappa \cos\theta
    \end{array}\right) \,,
\end{align}
where $\theta=\theta'+i\,\theta''$ is an arbitrary complex number and $\kappa=\pm 1$. Since we have only two right-handed neutrinos, we take $m_{\nu1}=0$ in $m^{\rm diag}_\nu$. We remark that such a parametrization can be used only when the relation for the light neutrino masses~\eqref{ss1} is valid, namely when $M_D\ll M_{R}$.

Clarifying the nature and the origin of neutrino mass is the biggest challenge of neutrino physics. From one side, neutrinoless double beta decay experiments can probe the Majorana nature of neutrinos; on the other side, experiments like SHiP have the purpose to shed light on neutrino mass generation mechanisms by detecting Heavy Neutral Leptons assumed in seesaw mechanisms. In SHiP right-handed neutrinos can be produced with a 400 GeV proton beam on a heavy fixed target. Then the HNLs decay in the detector acceptance, resulting in a detectable Standard Model final state. The decay rate is proportional to: 1) the mixings between the incoming and outcoming Standard Model neutrino with flavor $e,\mu,\tau$ and heavy sterile neutrinos; 2) the HNLs masses  $M_N$  with $N=1,2$. In principle one could have different cases depending on HNLs mass difference $\Delta M_{\rm HNL} \equiv  |M_2-M_1|$: 
\begin{itemize}
\item {\bf Degenerate:} $\Delta M_{\rm HNL} \ll m_\nu$;
\item {\bf Hierarchical:} $\Delta M_{\rm HNL} \gg m_\nu$.
\end{itemize}
In the mass range of sensitivity of SHiP, that is about $(0.6\div 6.0)$\,GeV, if the HNLs masses are such that $\Delta M_{\rm HNL} \lesssim 0.1~{\rm GeV}$, HNLs production-decay mechanism for the Standard Model flavor $\alpha=e,\mu,\tau$ state is a function of the mass $M_{\rm HNL}\equiv M_1\simeq M_2$ and the mixing
\begin{equation}\label{U2i}
U_{\alpha}^2 = \sum_{N=1}^{2} |U_{\alpha (N+3)}|^2 \,,
\end{equation}
where the sum is over the number of HNLs and $U_{\alpha (N+3)}$ is the unitary matrix that diagonalizes the mass matrix in eq.~\eqref{mnu}. In order to understand the order of magnitude of the coupling of HNLs with Standard Model inclusively, it is also useful to introduce the quantity
\begin{equation}\label{U2}
U^2\equiv \sum_{\alpha}U_{\alpha}^2\,.
\end{equation}
SHiP Collaboration considers $\nu$MSM as a benchmark model where $\Delta M_{\rm HNL}  \lesssim \mathcal{O}(1)$~GeV\footnote{Mass differences up to 0.1\,GeV and 2\,GeV are respectively considered in \cite{Eijima:2018qke} and \cite{Antusch:2017pkq}.} and therefore the Collaboration provides the sensitivity as a function of $M_{\rm HNL}\equiv M_1\simeq M_2$ and the mixing $U^2$  \cite{Graverini:2015dka} or $U_{\alpha}^2$  \cite{SHiP:2018xqw}. On the other hand, in case of $\Delta M_{\rm HNL} \gtrsim 0.1~{\rm GeV}$, the production-decay rate is not proportional to $U_{\alpha}^2 $ and a generalization requires a dedicated study (being a more complicate function of $M_N$ and $U_{\alpha N}$) that is beyond the scope of the present paper. Such a case is here also considered, but it is not of interest apart from an academic sense. In fact, the mixing parameter $U^2$ is of physical relevance in the amplitude for the processes analyzed by SHiP only for $\Delta M_{\rm HNL} \lesssim 0.1~{\rm GeV}$.

Using the parametrization given in eq.~\eqref{Rmat}, we find (as in \cite{Antusch:2017pkq})
\begin{align}\label{general}
U^2=\frac{m_{\nu 2}-m_{\nu 3}}{2}\left(\frac{1}{M_1}-\frac{1}{M_2}\right)\cos(2\theta')+\frac{m_{\nu 2}+m_{\nu 3}}{2}\left(\frac{1}{M_1}+\frac{1}{M_2}\right)\cosh(2\theta'')\,.
\end{align}
where $\theta'$ and $\theta''$ are respectively the real and imaginary part of the complex rotation angle. It is easy to see that the minimum value allowed for this quantity is obtained for $\theta''=0$ and is equal to
\begin{align}
U^2_{\rm min}\simeq \frac{m_{\nu 2}}{M_1}+\frac{m_{\nu 3}}{M_2} = \frac{m_{\nu 2}}{M_{\rm HNL}}+\frac{m_{\nu 3}}{M_{\rm HNL} + \Delta M_{\rm HNL}}\,.
\end{align}
which implies the existence of a lower bound in the plane $(M_{\rm HNL},U^2)$. 

It is also possible to ascertain the existence of an upper bound due to the fact that for high enough values of $\theta''$ the seesaw condition is not verified anymore.  This implies that the input values of the Casas-Ibarra parametrization, which are chosen in the experimental range, are not the same values which result from the diagonalization of the neutrino mass matrix. In fact, the Casas-Ibarra parametrization derives from~\eqref{ss1}, which is only valid in the seesaw regime in which the elements of $M_D$ are much smaller than the right-handed mass. In general, that relation receives corrections of higher orders in what we might call the seesaw ratio:
\begin{equation}
\zeta=\frac{{\rm max}_{ij}(M_{D_{ij}})}{M_{\rm HNL}}\,.
\end{equation}
One can prove that the seesaw corrections are roughly of order $\zeta^2$.\\
Then the real eigenvalues and eigenvectors of the full neutrino mass matrix will differ from the input values, due to these correction terms: in our numerical simulation, this will imply that, even if we take the input parameters to be well within the experimental confidence range, the difference between the output and input parameters might cause the output to be outside the experimental range and therefore excluded. As we saw before, the percentage correction to the input parameters in the output is roughly $\zeta^2$ (times numerical factors of order unity), and thus we expect that the lower limit will be obtained by taking a $\zeta^2$ of the order of magnitude of the smallest relative experimental uncertainties. From these uncertainties we can expect $\zeta$ to be of order $\zeta\sim 0.1$: comparison with experimental data shows that $\zeta\simeq 0.2$.\\
The Casas-Ibarra parametrization~\eqref{Cl1} shows that we need:
\begin{align}
\zeta\simeq \cosh\theta'' \sqrt{\frac{m_{\nu 3}}{M_{1}}} \,.
\end{align}
Since $M_{1}\gg m_{\nu 3}$, we can assume $\theta''$ to be large and substitute $\cosh\theta''\sim \frac{e^{\theta''}}{2}$. By inverting we then find:
\begin{align}\label{const}
e^{\theta''}\simeq 2\zeta \sqrt{\frac{M_{1}}{m_{\nu 3}}}\,.
\end{align}
If we substitute back into \eqref{general}, taking the case of $\cos\left(2\theta'\right)=1$, we find an upper limit given by:
\begin{align}
U^2_{max}\simeq\frac{m_{\nu3}-m_{\nu2}}{2}\left(\frac{1}{M_1}-\frac{1}{M_2}\right)+\zeta^2\frac{M_1}{m_{\nu3}}\left(m_{\nu2}+m_{\nu3}\right)\left(\frac{1}{M_1}+\frac{1}{M_2}\right)\simeq \zeta^2\frac{M_1}{m_{\nu3}}\left(m_{\nu2}+m_{\nu3}\right)\left(\frac{1}{M_1}+\frac{1}{M_2}\right)
\end{align}
where we have neglected the first term which, for the given values of the precision $\zeta$, is small by orders of magnitude with respect to the second. We will find that such high values of $U^2$ are already excluded from constraints coming from collider data and consistency with the double beta decay lifetime.

The analytical bounds obtained here are summarized in table~\ref{tabanalytical}.

\begin{table}[h]
\begin{center}
\begin{tabular}{|l|c|}
\hline
&  {\bf Bounds} \\
\hline
&\\
$U^2_{\rm max}$ &  $\zeta^2\frac{1}{m_{\nu3}}\left(m_{\nu2}+m_{\nu3}\right)\left(1+\frac{1}{1+\frac{\Delta M_{\rm HNL}}{M_{\rm HNL}}}\right)$ \\
&\\
\hline 
&\\
$U^2_{\rm min}$ & $\frac{m_{\nu 2}}{M_{\rm HNL}}+\frac{m_{\nu 3}}{M_{\rm HNL} + \Delta M_{\rm HNL}}$\\
&\\
\hline
\end{tabular}
\end{center}
\caption{Lower and upper bound for the quantity $U^2$ defined in eq.~\eqref{U2}. The parameter $\zeta$ is in the range $(0.01 \div 0.1)$ (see the text for details).}\label{tabanalytical}
\end{table}

It will be of interest to look at the behavior of $U^2$ as a function of $\Delta M_{\rm HNL}$ for a fixed value of $M_{\rm HNL}$. The lower limit is expected to be $\frac{m_{\nu 2}}{M_{\rm HNL}}+\frac{m_{\nu 3}}{M_{\rm HNL}+\Delta M_{\rm HNL}}$. If $M_{\rm HNL}$ is chosen to be much larger than $m_{\nu 3}$, as of course is natural to suppose, we find that
\begin{itemize}
\item for $m_{\nu 3}\ll \Delta M_{\rm HNL} \ll M_{\rm HNL}$:
\begin{equation}
U^2_{\rm min} (\Delta M_{\rm HNL} \ll M_{\rm HNL}) \sim \frac{m_{\nu 2}+m_{\nu 3}}{M_{\rm HNL}}
\end{equation}
\item for $\Delta M_{\rm HNL} \gg M_{\rm HNL}$:
\begin{equation}
U^2_{\rm min} (\Delta M_{\rm HNL} \gg M_{\rm HNL}) \sim  \frac{m_{\nu 2}}{M_{\rm HNL}}
\end{equation}
\end{itemize}
We therefore expect that in this latter regime the value of $U^2_{\rm min}$ falls by a ratio of 
\begin{align}
\frac{U^2_{\rm min} (\Delta M_{\rm HNL} \ll M_{\rm HNL})}{U^2_{\rm min} (\Delta M_{\rm HNL} \gg M_{\rm HNL})}=\frac{m_{\nu 2}}{m_{\nu 2}+m_{\nu 3}}.
\end{align}
From the previous discussion we can draw the conclusion that the physics behind the SHiP experiment critically depends on the value of the imaginary part of the complex angle of rotation $\theta''$. In fact, large values of $\theta''$ lead to extremely large values of the mixing. It might therefore be of interest to notice that, for extremely large values of $\theta''$, the matrix $M_D$ approaches a limit structure. In fact, both $\cosh\theta''$ and $\sinh\theta''$ can be approximated by $\frac{e^{\theta''-i\theta'}}{2}$. Therefore we have
\begin{align}\label{MDmb}
M_D\simeq \frac{e^{\theta''-i\theta'}}{2}\, U_{\rm PMNS} \cdot \left(\begin{array}{cc}
         0& 0  \\
         \sqrt{M_1 m_2} & -\sqrt{M_2 m_2} \\
         \sqrt{M_1 m_3} & \sqrt{M_2 m_3}
    \end{array}\right).
\end{align}

\section{Numerical results}

We will now present the numerical results of our analysis, which will be seen to confirm the analytical results previously derived and complement them with the limits coming from the double beta decay. Our analysis is performed with the following steps:
\begin{itemize}
\item
We select an arbitrary set of numerical values for $\theta_{12}^\nu$, $\theta_{13}^\nu$, $\theta_{23}^\nu$, $\Delta m^2_{sol}$, $\Delta m^2_{atm}$, $\delta$, $\alpha$, $\beta$, $\theta'$, $\theta''$, $\kappa$, $M_{\rm HNL}$, $\Delta M_{\rm HNL}$ according to the ranges reported in table~\ref{tab1};
\item  we generate a numerical $3\times 2$ Dirac neutrino mass matrix $M_D$ using the Casa-Ibarra parametrization eq.~\eqref{Cl1};
\item we diagonalize the numerical $5\times 5$ neutrino mass matrix $M_\nu$ using the methods described above;
\item we check that the eigenvalues and eigenvectors obtained fit neutrino oscillation data;
\item we take the arbitrary set of points only if all the above conditions are satisfied.
\end{itemize}
\begin{table}[t]
\begin{center}
\begin{tabular}{|c|c|}
\hline
Observable & Input range\\
\hline
$\sin^2\theta_{13}^\nu$ & $(1.90,2.39)\, 10^{-2}$\\
$\sin^2\theta_{23}^\nu$ & $(4.30,6.02)\, 10^{-1}$\\
$\sin^2\theta_{12}^\nu$ & $(2.65,3.46)\, 10^{-1}$\\
$\Delta m^2_{atm}$ & $(2.39,2,60)\,10^{-3}$~eV\\
$\Delta m^2_{sol}$ & $(6.92,7.91)\,10^{-5}$~eV\\
$\delta$ & $(0.83\,\pi,1.99\,\pi)$\\
\hline
\end{tabular}
\hskip5.mm
\begin{tabular}{|c|c|}
\hline
Parameter & Input range\\
\hline
$\alpha,\beta$ & $(0,2\pi)$\\
$\theta'$ & $(0,2\pi)$\\
$\theta''$ & $(0,30)$\\
$\kappa$ & $\pm 1$\\
$M_{\rm HNL}$ & $(0.1,10)$~GeV\\
$\Delta M_{\rm HNL}$ & $(10^{-20},10^{6})$~GeV\\
\hline
\end{tabular}
\end{center}
\caption{Observables and parameters ranges used in our numerical analysis. The measured neutrino oscillations parameters for normal ordering are taken from ref.~\cite{Capozzi:2018ubv}.}\label{tab1}
\end{table}

In the plots we also show the constraints on $U^2$ provided by colliders~\cite{Deppisch:2015qwa}, Big Bang Nucleosynthesis (BBN)~\cite{Canetti:2012kh} and the non-observation of the neutrinoless double beta decay~\cite{Agostini:2013mzu,Agostini:2018tnm,Shirai:2017jyz}. Regarding the latter, we require that the neutrinoless double beta $0\nu\beta\beta$ half-life $T^{0\nu}_{1/2}$ is larger then the experimental limit $T^{0\nu\,{\rm Ge}}_{1/2}=8.0\times 10^{25}$~s~\cite{Agostini:2013mzu,Agostini:2018tnm} and $T^{0\nu\,{\rm Xe}}_{1/2}=10.7\times 10^{25}$~s~\cite{Shirai:2017jyz} where
\begin{equation}\label{T}
\left[T^{0\nu}_{1/2}\right]^{-1} = \mathcal{A}
\left| \frac{m_p}{\langle p^2 \rangle}  \sum_{k=1}^3 U_{ek}^2 m_{\nu k} +m_p  \sum_{N=1}^{2}
\frac{U_{e(N+3)}^2 M_{N}}{\langle p^2 \rangle+M^2_N}
\right|^2  \,,
\end{equation}
where $m_p=938$MeV and the numerical values for $\mathcal{A}$ and $\langle p^2 \rangle$ are taken from different nuclear models in \cite{Faessler:2014kka}. For each model we have verified the consistency of our results with experimental data. The different parameters are reported in 
table~\ref{tab0nubb}.
\begin{table}[t]
\begin{center}
\begin{tabular}{|ll|c|c|c|c|}
\hline
 & & a& b& c& d\\
\hline 
$^{\,76}$Ge: & $\sqrt{\langle p^2 \rangle}$ [MeV] &  159& 163 & 190 & 193  \\
$^{136}$Xe: & $\sqrt{\langle p^2 \rangle}$ [MeV] &  178 & 183 & 208 & 211  \\
$^{\,76}$Ge: &  $\mathcal{A}~[10^{-10}{\rm yrs}^{-1}]$ &  2.55& 5.05 & 6.12 & 11.50  \\
$^{136}$Xe: & $\mathcal{A}~[10^{-10}{\rm yrs}^{-1}]$ &  4.41 & 8.74 & 10.40 & 19.70  \\
\hline
\end{tabular}
\end{center}
\caption{Numerical values for $\mathcal{A}$ and $\langle p^2 \rangle$ used in eq.~\eqref{T} for different nuclear models denoted by a,b,c,d as provided in reference \cite{Faessler:2014kka}.}\label{tab0nubb}
\end{table}

\begin{figure}[t]
\begin{center}
\includegraphics[width=0.325\textwidth]{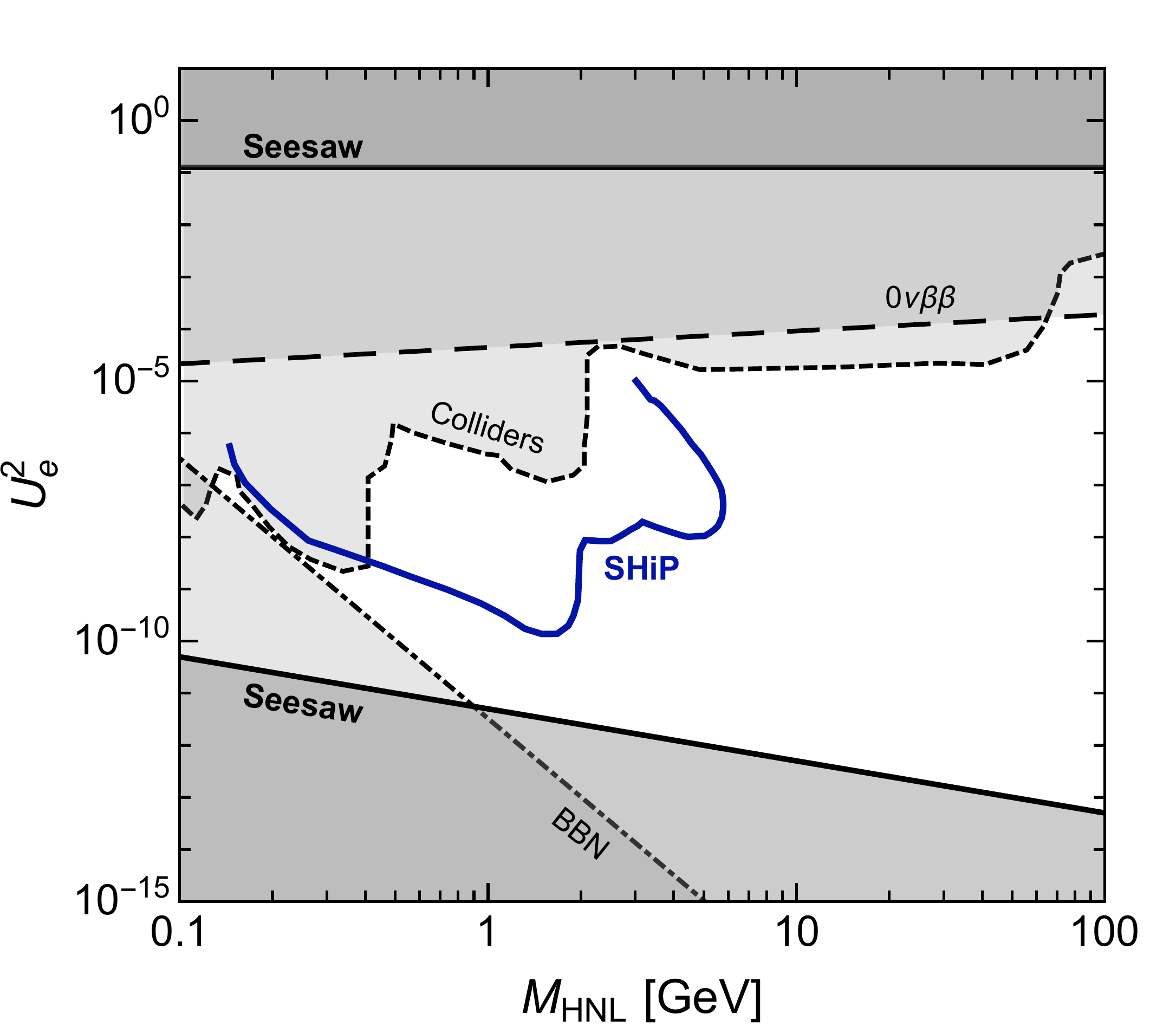}
\includegraphics[width=0.325\textwidth]{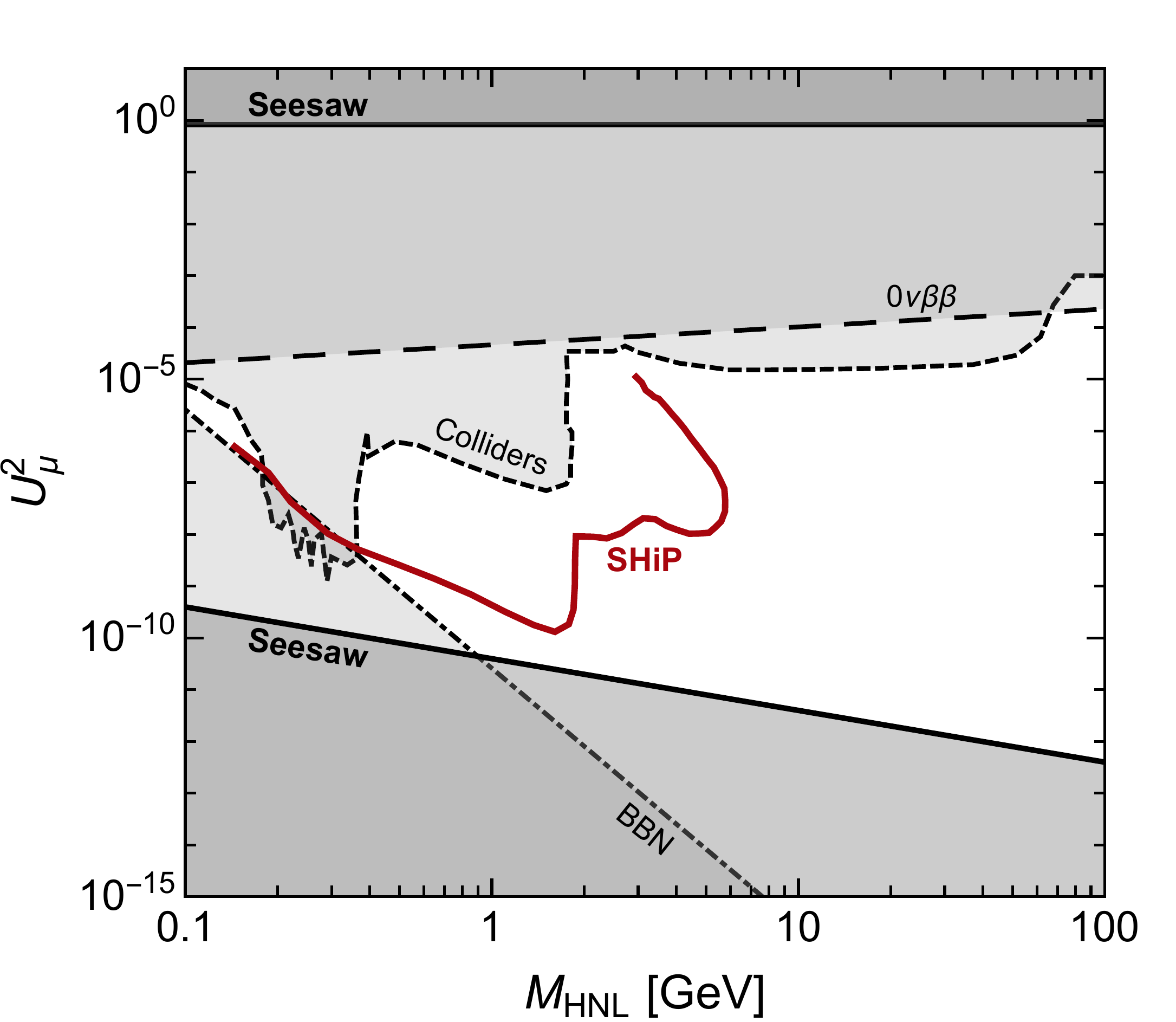}
\includegraphics[width=0.325\textwidth]{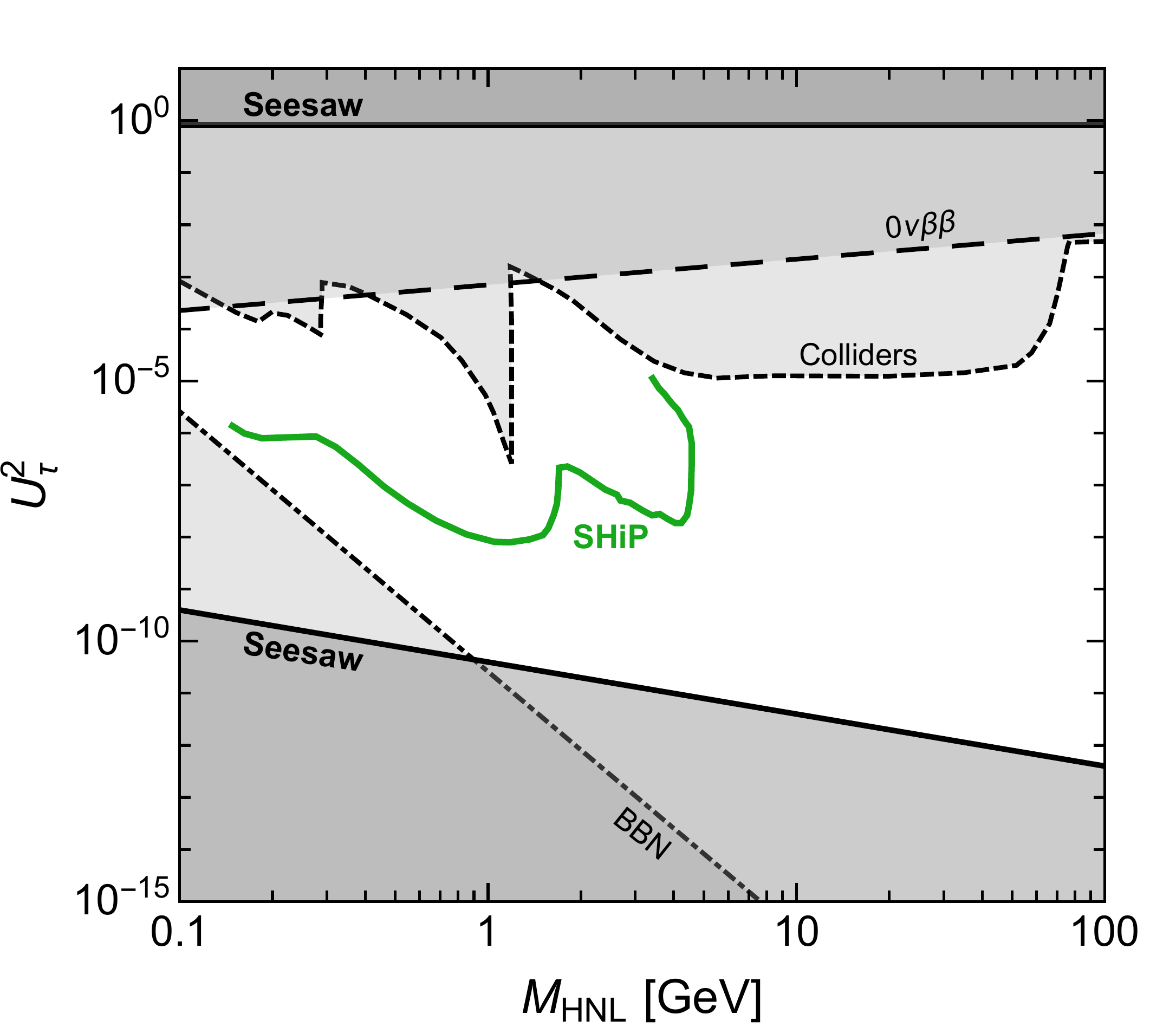}
\end{center}
\caption{\label{fig2G} $U_{e,\mu,\tau}^2$  vs $M_{\rm HNL}$ for 2HNL seesaw. SHiP sensitivity for different flavor parameters has been taken from \cite{SHiP:2018xqw} (colored lines). The two seesaw lines are the upper and lower limits predicted from the sole requirement of the seesaw condition. The current experimental limits (gray bands) are: the constraints by colliders data (short-dashed lines); the lower limits coming from BBN (dotted-dashed lines); the upper limits coming from the non-observation of neutrinoless double beta decay lifetime (long-dashed lines.}
\end{figure}

In  figure~\ref{fig2G} we report the three quantities $U_{e,\mu,\tau}^2$ as a function of $M_{\rm HNL}$ where all the bands shown have been obtained only numerically. In each panel we display the current experimental upper limits from colliders~\cite{Deppisch:2015qwa} and lower ones from BBN~\cite{Canetti:2012kh}, and the SHiP sensitivity~\cite{SHiP:2018xqw}. The two seesaw lines mark the region allowed from the seesaw mechanism; as the previous analytical discussion makes clear, these limits comes from the requirement of the seesaw condition in the case of hierarchical and degenerate right-handed neutrinos. The orange line describes instead the lower limit predicted solely in the case of hierarchical neutrinos, which roughly coincides with the upper limit of the degenerate case. In previous works (where the $\nu$MSM model had been analysed) this line had been considered as the effective lower limit for the region allowed from the seesaw. This is therefore the main difference between our work and previous works on the same subject.

\begin{figure}[t]
\begin{center}
\includegraphics[width=0.325\textwidth]{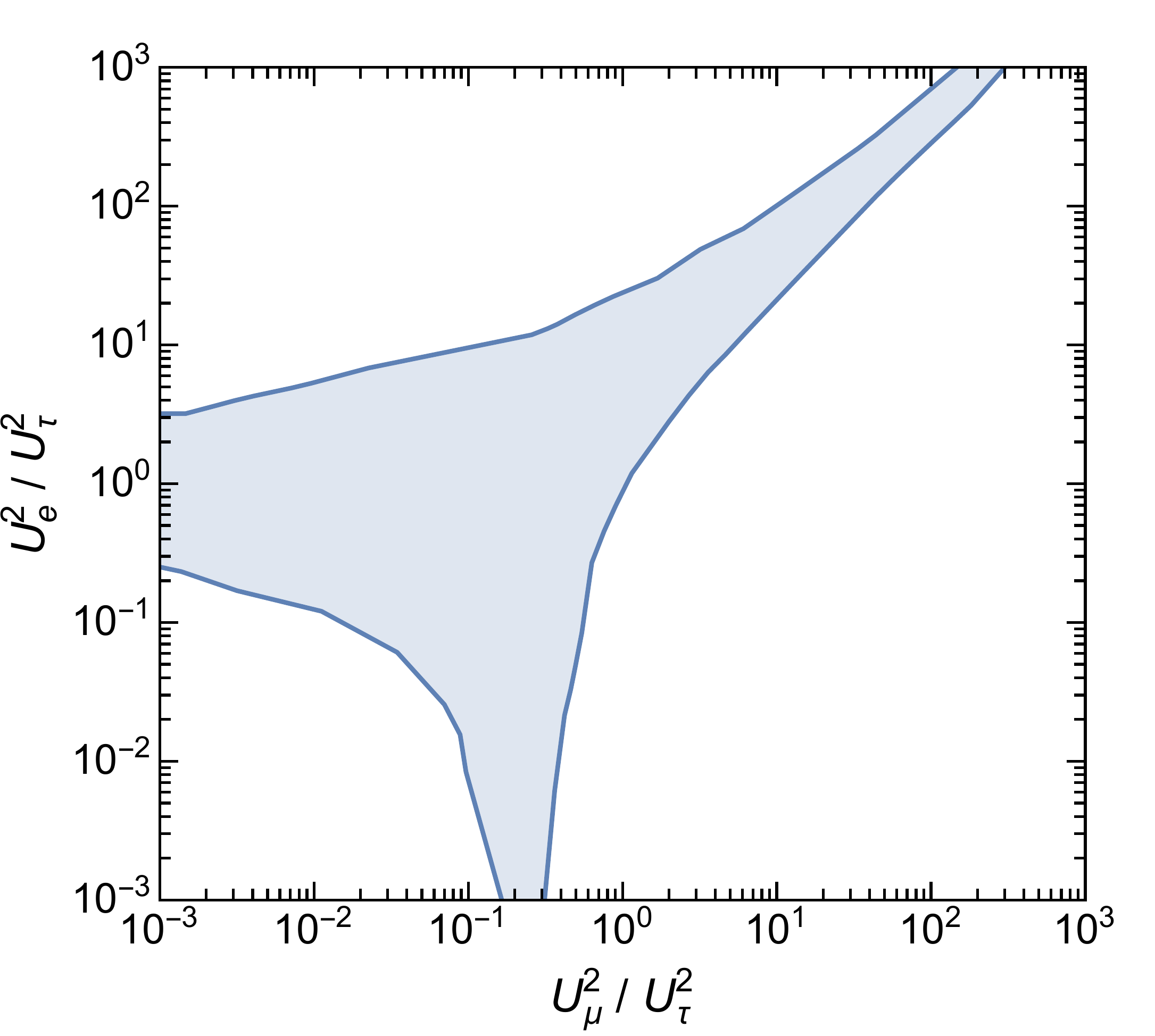}
\includegraphics[width=0.325\textwidth]{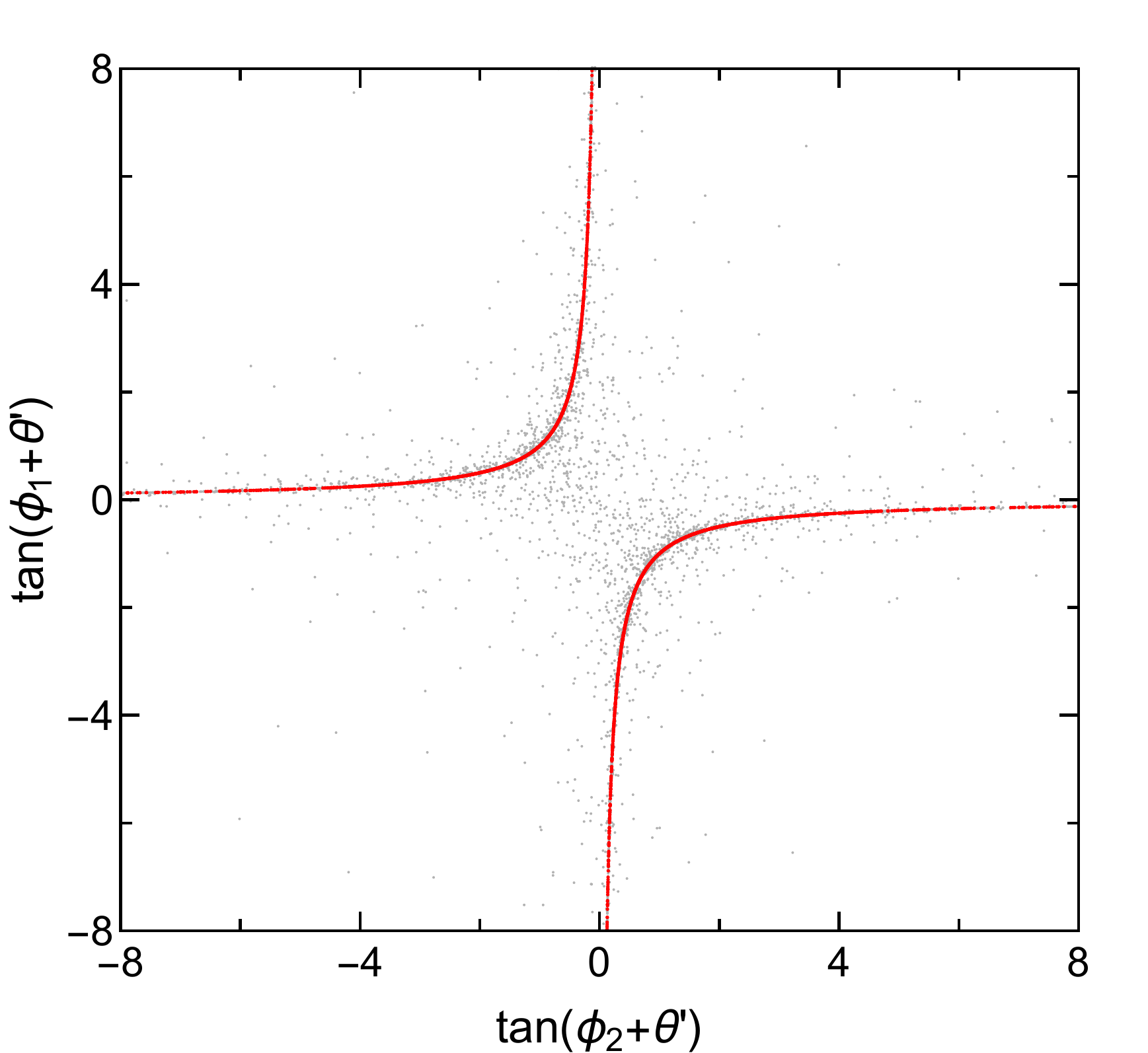}
\end{center}
\caption{\label{figU} {\it Left panel:} $U_{e,\mu,\tau}^2$ ratios obtained in our numerical model-independent analysis. {\it Right panel:} Phases of the element $M_{D11}$ and $M_{D12}$, where the red points correspond to $U^2\geq 10^{-7}$, while the gray points have $U^2\leq10^{-7}$.}
\end{figure}

Finally, in the left panel of figure~\ref{figU} we show our model-independent results for the allowed flavor ratio $\left(U_{e}^2: U_{\mu}^2: U_{\tau}^2\right)$. Moreover, in the right panel of the same figure we depict the presents of relations that are satisfied by the element of the full complex PMNS matrix. Indeed, by substituting the full complex PMNS matrix into eq.~\eqref{MDmb}, one can obtain a number of relations which are verified in the limit of large $\theta''$ in which that equation was obtained. For example, if $\phi_1$ and $\phi_2$ are respectively the phases of the element $M_{D11}$ and $M_{D12}$, then they obey
\begin{equation}
\tan\left(\phi_1+\theta'\right) \tan\left(\phi_2+\theta'\right)=-1\,.
\label{eq:rel}
\end{equation}
This relation has been verified to hold in the limit of large values of $\theta''$, where the red points correspond to where the quantity $U^2$ is greater than a threshold value of $10^{-7}$. It is therefore evident that, in the hierarchical case, the greater $U^2$, the better is the relation of eq.~\eqref{eq:rel} verified, due to the large values of $\theta''$ necessary to obtain an enhancement in $U^2$. We observe that such a correlation is not necessary in case of low-energy seesaw like the inverse one~\cite{Mohapatra:1986aw,Mohapatra:1986bd} where one can have very low masses for HNLs but with order one Yukawa couplings in a natural way differently from type-I seesaw, where for HNLs masses in the range $(0.1\div10)$~GeV tiny Yukawa couplings $M_D/v$ (where $v$ is the electroweak v.e.v.) of order of $10^{-8}\div 10^{-6}$ are necessary. 

\vskip5.mm

\section{Conclusions}

We study analytically and numerically a type-I seesaw neutrino mass mechanism with two heavy neutral leptons (right-handed neutrinos).
This kind of seesaw mechanism will be investigated at SHiP through the measurement of the active-sterile mixing angle $U^2$. Contrary to a naive expectation, which would presume $U^2$ to be of order $10^{-11}$ for HNLs masses of order of $\sim 1$GeV,  the seesaw is actually able to predict mixing angles in the range $(10^{-11},10^{-1})$, confirming previous results. Of course, this seesaw admitted region is further constrained from data coming from colliders, BBN and neutrinoless double beta decay. We find that, in order to obtain such extreme values for $U^2$, some specific relations between the phases of the Dirac coupling are to be satisfied.\\ In our numerical analysis we have also included 
neutrinoless double beta decay experimental bounds that do not provide any restriction for SHiP.

\end{document}